\documentclass[aps,pre,amsmath,amssymb,twocolumn]{revtex4}
\usepackage{graphicx}
\usepackage{amssymb}
\usepackage{amsmath}
\usepackage{epstopdf}
\newcommand{\be}{\begin{equation}}\newcommand{\ee}{\end{equation}}
\newcommand{\ba}{\begin{array}{l}}\newcommand{\ea}{\end{array}}
\newcommand{\baa}{\begin{eqnarray}}\newcommand{\eaa}{\end{eqnarray}}
\newcommand{\lab}[1]{\label{#1}}\newcommand{\re}[1]{(\ref{#1})}
\newcommand{\ci}[1]{\cite{#1}}
\usepackage[utf8]{inputenc}
\usepackage{makeidx}
\usepackage{fourier}

\begin{document}

\title {Soliton generation in optical fiber networks}
\author{K.K. Sabirov$^{a,b}$,  M. E. Akramov$^c$, Sh. R. Otajonov$^d$ and D.U. Matrasulov$^a$}
\affiliation{ $^a$ Turin Polytechnic University in Tashkent, 17
Niyazov Str.,
100095,  Tashkent, Uzbekistan\\
$^b$ Tashkent University of Information Technologies, 108 Amir Temur Str., 100200, Tashkent Uzbekistan\\
$^c$National University of Uzbekistan, Vuzgorodok, Tashkent
100174,Uzbekistan\\
$^d$Physical-Technical Institute, Uzbekistan Academy of Sciences,
2-b, Bodomzor str., 100084, Tashkent, Uzbekistan}

\begin{abstract}
We consider the problem of soliton generation in branched optical
fibers. A model based on the nonlinear Schrodinger equation on
metric graphs is proposed. Number of generated solitons is
computed for different branching topologies considering different
initial pulse profiles. Experimental realization of the model is
discussed.
\end{abstract}
\maketitle
\section{Introduction}
Optical solitons attracted much attention due to their potential
applications in optoelectronics and information technologies. The
idea of using optical solitons as carriers of information in
high-speed communication systems was first proposed in the
pioneering paper by Hasegawa and Tappert \ci{Hasegawa}. Later due
to the advances made in fiber technology it became possible to
realize  optical solitons experimentally in different versions
(bright, dark, etc) \ci{Fiber1}-\ci{Kivsharbook}. This fact caused
great interest to finding the soliton solutions of governing
nonlinear wave equations, such as nonlinear Sch\"odinger equations
with different nonlinearities. An important problem in the context
of optical solitons is the problem of soliton generation in
optical media. Mathematically, such problem is reduced to the
initial value (Cauchy) problem for nonlinear Sch\"odinger
equation, which allows to find soliton solution and number of
generated solitons using given initial condition. For optical
fibers such problem was studied in the
Refs.\ci{Burzlaff}-\ci{Zhong1}. In \ci{Burzlaff} an effective
method for finding number of solitons generated in optical fibers
was proposed. Later, it was extended for some other initial
conditions \ci{Kivshar}.  Strict mathematical treatment of soliton
generation on a half line as initial-boundary-value problem was
considered presented \ci{Fokas}.  Soliton generation in optical
fibers for a dual-frequency input was studied in \ci{Panoiu}.  In
\ci{Skryabin} a theory of the generation of new spectral
components in optical fibers pumped with a solitonic pulse and a
weak continuous wave was proposed and  the wave number matching
conditions for this process was  derived. In \ci{Nishizawa}
characteristics of wavelength-tunable femtosecond  soliton pulse
generation using optical fibers in a negative dispersion region
are studied experimentally and theoretically using the  extended
nonlinear Schr\"odinger equation, in which the wavelength
dependence of parameters is considered. A comprehensive analysis
of the generation of optical solitons in a monomode optical fibre
from a superposition of soliton-like optical pulses at different
frequencies in \ci{Panoiu1}, where it is found that there exists a
critical frequency separation above which wavelength-division
multiplexing with solitons is feasible. Soliton generation and
their instability are investigated in a system of two
parallel-coupled fibers, with a pumped (active) nonlinear
dispersive core and a lossy (passive) linear one in \ci{Malomed1}.
A theory of the generation of new spectral components in optical
fibers pumped with a solitonic pulse have been studied.
Bright-gap-soliton generation in finite optical lattices was
discussed in \ci{Carusotto}. Despite the fact that certain
progress is made on theoretical and experimental study of soliton
generation in optical fibers, all the studies are restricted by
considering long, unbranched fibers. However, branched fibers are
more attractive from the viewpoint of practical applications, as
in may cases information-communication systems use optical fiber
networks. Modeling of soliton generation and dynamics in optical
fiber networks requires solving of nonlinear Sch\"odinger equation
on metric graphs.\\
We note that soliton dynamics described by integrable nonlinear
wave equations attracted much attention during past decade
\ci{Zarif} -\ci{Karim2018}. In \ci{Zarif} nonlinear Sch\"odinger
equation on metric graphs is studied and  condition for
integrability is derived in the form of a sum rule for
nonlinearity coefficients. In \ci{zar2011} such study is extended
to Ablowitz-Laddik equation. Stationary Sch\"odinger equation on
metric graphs and standing wave soliton in networks are studied in
\ci{Adami,noja,Karim2013,Adami16,Adami17}. Integrable sine-Gordon
equation on metric graphs is studied in
\ci{caputo14,Our1,Karim2018}. Linear and nonlinear systems of PDE
on metric graphs  are considered in \ci{Bolte,KarimBdG,KarimNLDE}.

In this paper we consider the problem of soliton generation in
branched optical fibers, or, optical fiber networks described in
terms of the initial value problem for nonlinear Sch\"odinger
equation on metric graphs. For different given initial conditions,
we derive number of solitons generated by considering different
network topologies. Unlike linear optical fibers,  pulse
generation and soliton dynamics in  fiber networks strongly depend
on the topology of latter. Propagating in such network optical
soliton undergo to scattering and transmission through the network
branching points that may cause additional effects such as
interaction of incoming and scattered solitons, radiation,
collisions, etc. Therefore effective transmission of information
through the optical fiber networks requires proper tuning both the
system architecture and initial pulse profiles. Depending on which
branch, or vertex the initial pulse located, the number of
solitons and their dynamics can be different. This fact provides
powerful tool for tuning of the optical fiber network architecture
and optimization of signal and information transmission.  This
paper is organized as follows. In the next section we briefly
recall treatment of the problem of soliton generation for linear
(unbrached) optical fibers. In section III we give formulation of
the problem and its solution for star branched (Y-junction)
optical fibers. Section IV extends the study for other network
topologies, modeled by loop and tree graphs. Section V presents
some concluding remarks.

\begin{figure}[h]
\centering
\includegraphics[width=80mm]{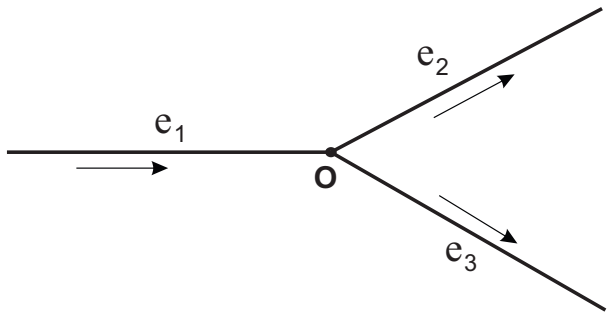}
\caption{Basic star graph} \label{f2}
\end{figure}

\section{Soliton generation in linear optical fibers}

Let us first, following the Refs.\ci{Burzlaff,Kivshar},  recall
solution of the problem for linear, i.e. unbranched optical
fibers. The governing equation for the pulse generation and
evolution in optical fibers is the following nonlinear
Sch\"odinger equation \be i\frac{\partial \psi}{\partial
t}+\frac{1}{2}\frac{\partial^2 \psi}{\partial
x^2}+|\psi|^2\psi=0,\lab{nls01} \ee where $\psi$ is the normalized
complex amplitude of the pulse envelope. The problem of soliton
generation in optical fibers is reduced to the Cauchy problem for
Eq.\re{nls01}. Such problem can be solved, e.g., using inverse
scattering method \ci{Burzlaff,Kivshar,Panoiu}. In \ci{Burzlaff}
it was solved for the initial conditions given by
$\psi(x,0)=-iq(x)$, with
\begin{eqnarray}
q(x)=
\begin{cases}
0, & \text{for}\;\;\;  |x|>\frac{1}{2}a \\
b, & \text{for}\;\;\;  |x|\leq\frac{1}{2}a
\end{cases}\quad\quad\quad b>0.
\end{eqnarray}

The evolution of the wave function upon generation of the soliton
can be obtained via solving the following eigenvalue problem \be
Au =\lambda u, \ee where \be A =
\left(\begin{array}{cc}i\frac{d}{dx}&\psi(x,0)\\-\psi^*(x,0)&-i\frac{d}{dx}\end{array}\right).\lab{current}\ee
\begin{figure}[th!]
\includegraphics[width=80mm]{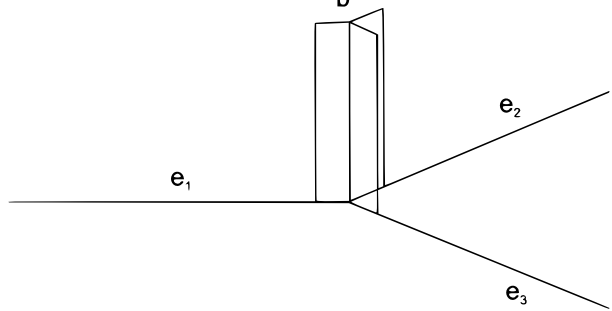}
\caption{Initial pulse profile in the star branched optical
fiber.} \label{psg}
\end{figure}

Each discrete eigenvalue $\lambda = \xi + i\eta $ with $L^2-$
integrable eigenfunction corresponds to the generated soliton with
the amplitude $2\eta$ moving with the velocity $2\xi$. It was
shown in \ci{Burzlaff} that number of generated solitons is given
by expression \be N = <\frac{1}{2} +\frac{F}{\pi}> = <\frac{1}{2}
+\frac{ab}{\pi}>,\lab{number}\ee where $F
=\int_{-\infty}^{\infty}|\psi(x,0)|dx$ and $<...>$ denotes the
integer smaller than the argument. Similar result for the number
of solitons was obtained in the Ref.\ci{Burzlaff} for the initial
condition given by
$$
q(x) =\beta \text{exp}(-\alpha |x|), \;\; \alpha, \beta > 0.
$$

Later, Kivshar   considered the problem of soliton generation for
super Gaussian initial pulse and showed that Eq.\re{number} is
general formula for arbitrary initial profile \ci{Kivshar}. More
detailed treatment of the problem of  soliton generation in
optical fibers was presented in \ci{Panoiu}. In particular, the
authors of \ci{Panoiu} analyzed  scenarios for soliton generation
in an ideal fiber for an input that consists of either two
in-phase or out-of-phase solitonlike optical pulses at different
frequencies by considering symmetric initial input pulse given by
$$
\psi(x,0) = \text{sech}(x)[e^{i\omega x} + e^{-i\omega x}],
$$
and asymmetric pulse given by
$$
\psi(x,0) = i\text{sech}(x)[e^{i\omega x} - e^{-i\omega x}]
$$
with the soliton solutions, respectively given as
\begin{widetext}
$$
\psi(x,t)=\xi\eta e^{\phi(\frac{t}{2})}\frac{e^{i\xi
x}\cosh\left[\eta(x+\xi t)+i\varphi\right]+e^{-i\xi
x}\cosh\left[\eta(x-\xi t)-i\varphi\right]}{\xi^2\cosh\eta(x+\xi
t)\cosh\eta(x-\xi t)+\eta^2\sin\xi(x+i\eta t)\sin\xi(x-i\eta t)},
$$
and
\begin{equation}
\psi(x,t)=-i\xi\eta e^{\phi(\frac{t}{2})}\frac{e^{i\xi
x}\cosh\left[\eta(x+\xi t)+i\varphi\right]-e^{-i\xi
x}\cosh\left[\eta(x-\xi t)-i\varphi\right]}{\xi^2\cosh\eta(x+\xi
t)\cosh\eta(x-\xi t)+\eta^2\cos\xi(x+i\eta t)\cos\xi(x-i\eta
t)},\label{linesol2}
\end{equation}
\end{widetext}
where
$\phi(t)=-i(\xi^2-\eta^2)t+\alpha,\,\alpha=\ln|\lambda_0|,\,\tan\varphi=\frac{\eta}{\xi}$.

In the next section we extend these studies to the case of
branched optical fibers, i.e. fiber networks.

\begin{figure}[h]
\centering
\includegraphics[width=80mm]{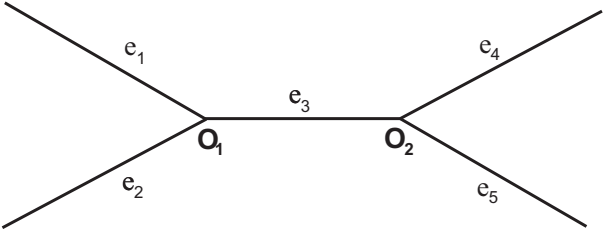}
\caption{Sketch for the H-graph} \label{f3}
\end{figure}

\section{Description of the model for Star shaped network}

Soliton dynamics in networks has attracted much attention during
past decade.  Convenient approach to describe such system is
modeling in terms of nonlinear wave equation on a metric graph.
The early treatment of the nonlinear Schrodinger equation on
metric graphs dates back to the Ref.~\cite{Zarif}, where soliton
solutions of the NLSE on metric graphs was obtained and
integrability of the problem under certain constraints was shown
by proving the existence of an infinite number of conserving
quantities.

Here we briefly recall the problem of NLSE in metric star graph
following the Ref.~\cite{Zarif}. Consider the star graph with
three bonds $e_j$ (see, Fig. 1), for which a coordinate $x_j$ is
assigned. Choosing the origin of coordinates at the vertex, 0 for
bond $e_1$ we put $x_1\in (-\infty,0]$ and for $e_{2,3}$ we fix
$x_{2,3}\in [0,+\infty)$. In what follows, we use the shorthand
notation $\Psi_j(x)$ for $\Psi_j(x_j)$ where $x$ is the coordinate
on the bond $j$ to which the component $\Psi_j$ refers. The
nonlinear Schr\"odinger equation on each bond $e_j$ of such graph
can be written as \ci{Zarif}\be i\frac{\partial \psi_j}{\partial
t}+\frac{\partial^2 \psi_j}{\partial
x^2}+\beta_j|\psi_j|^2\psi_j=0.\lab{nls02} \ee

To solve Eq. \eqref{nls02} one needs to impose the boundary
conditions at the branching point. This can be derived from the
fundamental physical laws, such as norm and energy conservation,
which are given as \ci{Zarif}\be
  \frac{dN}{dt} = 0, \quad\frac{dE}{dt} =0, \label{conserv}
\ee where
\begin{equation*}
N(t) =\int\limits_{-\infty}^{0}|\psi_1|^2\,dx
+\int\limits_{0}^{\infty}|\psi_2|^2\,dx
+\int\limits_{0}^{\infty}|\psi_3|^2\,dx
\end{equation*}
and
\begin{equation*}
E =E_1+E_2+E_3,
\end{equation*}
with
\begin{equation*}
E_k =\int\limits_{e_k}\biggl[\left|\frac{\partial\psi_k}{\partial
x}\right|^2 -\frac{\beta_k}{2}|\psi_k|^4\biggr]\,dx.
\end{equation*}
 As it was shown in the Ref. \ci{Zarif}, the conservation laws
Eq.~\eqref{conserv} lead to the following vertex conditions \be
  \sqrt{\beta_1} \psi_1(0,t)=\sqrt{\beta_2} \psi_2(0,t)=\sqrt{\beta_3} \psi_3(0,t),
\label{vbc1} \ee and generalized Kirchhoff rules \be
\frac{1}{\sqrt{\beta_1}} \frac{d\psi_1}{dx}|_{x=0} =
\frac{1}{\sqrt{\beta_2}} \frac{d\psi_2}{dx}|_{x=0} +
\frac{1}{\sqrt{\beta_3}} \frac{d\psi_3}{dx}|_{x=0}, \label{vbc2}
\ee where $\beta_j$ are nonzero real constants. The asymptotic
conditions for Eq.~\eqref{nls01} are imposed as \be
  \lim_{|x|\to +\infty} \psi_j=0. \label{asymp}
\ee

The single soliton solutions of Eq.~\eqref{nls01} fulfilling the
vertex boundary conditions \eqref{vbc1}, \eqref{vbc2} and the
asymptotic condition, \eqref{asymp} can be written as \cite{Zarif}
\be
  \psi_j(x,t) = a\,\sqrt{\frac{2}{\beta_j}}\,\frac{\exp\bigl[i\frac{vx}{2}-
i(\frac{v^2}{4}-a^2)t\bigr]}{\cosh[a(x-l -vt)]}, \ee where the
parameters $\beta_j$ fulfill the sum rule \ci{Zarif}\be
  \frac{1}{\beta_1}=
  \frac{1}{\beta_2}+\frac{1}{\beta_3}.\label{sumrule}
\ee Here $v$, $l$ and $a$ are bond-independent parameters
characterizing velocity, initial center of mass and amplitude of a
soliton, respectively.

Consider branched optical fiber having the form of the Y-junction.
Such system can be considered as basic star graph presented in
 Fig. 1. Then the problem of generation
of soliton and its propagation can be modeled in terms of the
Cauchy problem for nonlinear Sch\"odinger equation on a basic star
graph, which is given by Eq.\re{nls02} for which the following
initial condition is imposed:
$$
\psi_j(x,0) = -i\sqrt{\frac{2}{\beta_j}}q_j(x,).
$$
Here $\psi_j$ is the normalized complex amplitude of the pulse
envelope on $j$th bond (branch) of the graph and $q_j(x)$ is the
initial profile of the amplitude. To solve this equation, one
needs to impose the boundary conditions at the branching point
(vertex) of the graph and determine the asymptotic of the wave
function at the branch ends. These can be written in the form of
Eqs. \re{vbc1}, \re{vbc2} and \re{asymp}.

\begin{figure}[h]
\centering
\includegraphics[width=80mm]{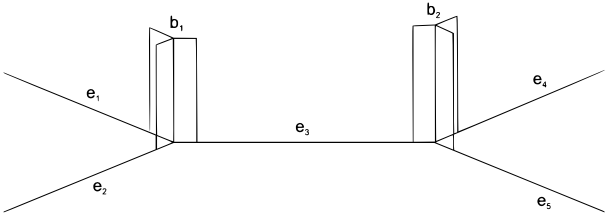}
\caption{Initial pulse profile on the optical fiber H-shaped
network} \label{f3}
\end{figure}

Here we consider the problem of soliton generation for Y-junction
of the optical fiber for the initial pulse profile given as (see,
Fig. \ref{f2}) \be q_1(x)=\begin{cases}0, & x<-\frac{1}{2}a \\ b,
& -\frac{1}{2}a\leq x\leq 0  \end{cases} \lab{profile00}\ee \be
q_{2,3}=\begin{cases}0, & x>\frac{1}{2}a \\ b, & 0\leq x\leq
\frac{1}{2}a  \end{cases} \lab{profile01} \ee Such initial profile
implies that soliton is generated around the branching point on
each branch.  Using the same method as that for linear optical
fiber, we can compute the number of generated solitons, $N$ for
such profile:
\begin{eqnarray}
N=\left<\frac{1}{2}+\frac{F}{\pi}\right>, \lab{number1}
\end{eqnarray}
where
\begin{eqnarray}
F=\sum_{j=1}^3\int_{e_j}|\psi_j(x,0)|dx=\frac{ab}{2}\left[
\sqrt{\frac{2}{\beta_1}}+\sqrt{\frac{2}{\beta_2}}+\sqrt{\frac{2}{\beta_3}}
\right]. \lab{number2}
\end{eqnarray}

We assume that the sum rule is \re{sumrule} is fulfilled, i.e. the
problem is integrable. Difference between Eqs.\re{number} and
\re{number1} comes from the constant factor
$$(\sqrt{2\beta_1^{-1}}+\sqrt{2\beta_2^{-1}}+
\sqrt{2\beta_3^{-1}}).$$  This allows tuning the soliton number
and dynamics using different choices of the set
$\beta_j,\;(j=1,2,3).$ In addition, for simplicity, the above
initial pulse profiles in Eqs.\re{profile00} and \re{profile01}
are given at the vertex and have the same widths, $a$ and heights,
$b$. However, in general case one can choose different widths and
heights for different bonds. This also provides additional tool
for tuning of the soliton number and dynamics.

Another initial pulse profile, for which the soliton number and
solutions in a Y-junction fiber can be explicitly obtained is
given by
\begin{equation}
\psi_j(x,0)=\sqrt{\frac{2}{\beta_j}}\text{sech}(x)\left[e^{i\left(\omega
x+\frac{\theta}{2}\right)}+e^{-i\left(\omega
x+\frac{\theta}{2}\right)}\right],\label{incond1}
\end{equation}
where $2\omega$ and $\theta$ are the frequency detuning and the
phase difference between the two solitons, correspondingly. The
two-soliton solution of the problem given by Eqs. \re{nls02},
\re{vbc1} and \re{vbc2}can be written as
\begin{widetext}
\begin{equation}
\psi_j(x,t)=\sqrt{\frac{2}{\beta_j}} \xi\eta
e^{\phi(\frac{t}{2})}\frac{e^{i\xi x}\cosh\left[\eta(x+\xi
t)+i\varphi\right]+e^{-i\xi x}\cosh\left[\eta(x-\xi
t)-i\varphi\right]}{\xi^2\cosh\eta(x+\xi t)\cosh\eta(x-\xi
t)+\eta^2\sin\xi(x+i\eta t)\sin\xi(x-i\eta t)},\label{starsol1}
\end{equation}
\end{widetext}
which is valid under the constraint:
\begin{equation}
\frac{1}{\beta_1}=\frac{1}{\beta_2}+\frac{1}{\beta_3},\label{const1}
\end{equation}
Corresponding  soliton number is given  by Eq.\re{number1}, where
the quantity $F$ can be written as
$$
F=2\pi\left(\sqrt{\frac{2}{\beta_1}}+\sqrt{\frac{2}{\beta_2}}+\sqrt{\frac{2}{\beta_3}}\right)\text{sech}\left(\frac{\pi\omega}{2}\right)\cos\left(\frac{\theta}{2}\right)
$$
\begin{figure}[h]
\centering
\includegraphics[width=80mm]{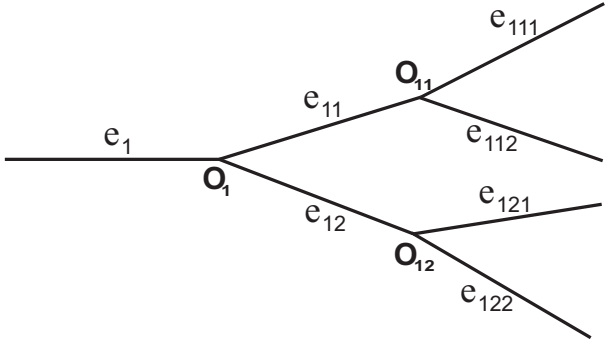}
\caption{Sketch for tree graph} \label{f4}
\end{figure}

\section{Other network topologies}

The above treatment of the problem for soliton generation in
optical fiber networks can be extended to the case of more
complicated topologies. Here we demonstrate this for so-called
$H-$graph and tree graph. For H-graph, presented in Fig.3 the
coordinates are defines as
$x_{1,2}\in(-\infty;0],\,x_3\in[0;L],\,x_{4,5}\in[0;+\infty),$
where $L$ is the length of bond $e_3$, i.e. the distance between
two vertices.

For NLS equation \re{nls02} and the vertex boundary conditions
given by
\begin{eqnarray}
  \sqrt{\beta_1} \psi_1(0,t)=\sqrt{\beta_2} \psi_2(0,t)=\sqrt{\beta_3} \psi_3(0,t),\nonumber\\
  \sqrt{\beta_3} \psi_3(L,t)=\sqrt{\beta_4} \psi_4(0,t)=\sqrt{\beta_5} \psi_5(0,t),
  \lab{vbc3}
\end{eqnarray}
\begin{eqnarray} \frac{1}{\sqrt{\beta_1}}
\frac{d\psi_1}{dx}|_{x=0} + \frac{1}{\sqrt{\beta_2}}
\frac{d\psi_2}{dx}|_{x=0} = \frac{1}{\sqrt{\beta_3}}
\frac{d\psi_3}{dx}|_{x=0},\nonumber\\
\frac{1}{\sqrt{\beta_3}} \frac{d\psi_3}{dx}|_{x=L} =
\frac{1}{\sqrt{\beta_4}} \frac{d\psi_4}{dx}|_{x=0} +
\frac{1}{\sqrt{\beta_5}} \frac{d\psi_5}{dx}|_{x=0}.\lab{vbc4}
\end{eqnarray}

We consider the following initial conditions:
$$\psi_j(x,0)=-i\sqrt{\frac{2}{\beta_j}}q_j(x)$$ where the initial
pulse profiles are given by (see, Fig.4)
\begin{eqnarray}
q_{1,2}(x)=
\begin{cases}
0, & -\infty< x<-\frac{1}{2}a \\
b_1, & -\frac{1}{2}a\leq x\leq 0
\end{cases}
\end{eqnarray}
\begin{eqnarray}
q_3(x)=
\begin{cases}
b_1, & 0\leq x \leq \frac{1}{2}a \\
0, & \frac{1}{2}a<x<L-\frac{1}{2}a\\
b_2, & L-\frac{1}{2}a \leq x \leq L
\end{cases}
\end{eqnarray}
\begin{eqnarray}
q_{4,5}(x)=
\begin{cases}
b_2, & 0\leq x \leq \frac{1}{2}a \\
0, & \frac{1}{2}a<x<\infty
\end{cases}
\end{eqnarray}

\begin{figure}[h]
\centering
\includegraphics[width=70mm]{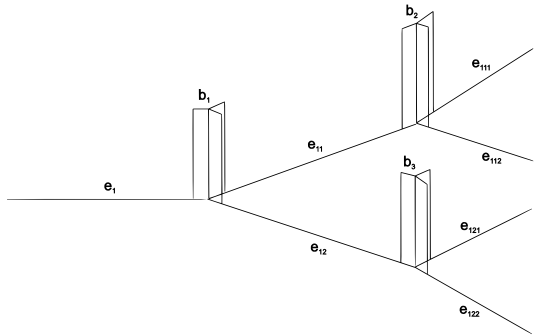}
\caption{Initial pulse profile on the tree-shaped  optical fiber
network} \label{f4}
\end{figure}

Then the number of generated optical solitons in such system we
have explicit expression:
\begin{eqnarray}
N=\left< \frac{1}{2}+\frac{F_1+F_2}{\pi} \right>,
\end{eqnarray}
where
\begin{eqnarray}
F_1=\frac{ab_1}{2}\left[
\sqrt{\frac{2}{\beta_1}}+\sqrt{\frac{2}{\beta_2}}+\sqrt{\frac{2}{\beta_3}}
\right],
\end{eqnarray}
and
\begin{eqnarray}
F_2=\frac{ab_2}{2}\left[
\sqrt{\frac{2}{\beta_3}}+\sqrt{\frac{2}{\beta_4}}+\sqrt{\frac{2}{\beta_5}}
\right].
\end{eqnarray}

Similarly, one can find number of generated solitons for the tree
graph, presented in Fig. \ref{f4}. The vertex boundary conditions
for such graph are given by
\begin{eqnarray}
  \sqrt{\beta_1} \psi_1(0,t)=\sqrt{\beta_2} \psi_2(0,t)=\sqrt{\beta_3} \psi_3(0,t),\nonumber\\
  \sqrt{\beta_{1i}} \psi_{1i}(L_{1i},t)=\sqrt{\beta_{1ij}}
  \psi_{1ij}(0,t),\,i,j=1,2,
  \lab{vbc3}
\end{eqnarray}
and
\begin{eqnarray} \frac{1}{\sqrt{\beta_1}}
\frac{d\psi_1}{dx}|_{x=0} = \frac{1}{\sqrt{\beta_2}}
\frac{d\psi_2}{dx}|_{x=0} + \frac{1}{\sqrt{\beta_3}}
\frac{d\psi_3}{dx}|_{x=0},\nonumber\\
\frac{1}{\sqrt{\beta_{1i}}} \frac{d\psi_{1i}}{dx}|_{x=L_{1i}} =
\frac{1}{\sqrt{\beta_{1i1}}} \frac{d\psi_{1i1}}{dx}|_{x=0} +
\frac{1}{\sqrt{\beta_{1i2}}}
\frac{d\psi_{1i2}}{dx}|_{x=0},\,i=1,2.\lab{vbc4}
\end{eqnarray}

Furthermore, we choose the initial pulse profile at each vertex
($\psi_e(x,0)=-i\sqrt{\frac{2}{\beta_e}}q_e(x)$) in the forms
\begin{eqnarray}
q_1(x)=
\begin{cases}
0, & -\infty< x<-\frac{1}{2}a \\
b_1, & -\frac{1}{2}a\leq x\leq 0
\end{cases}
\end{eqnarray}
\begin{eqnarray}
q_{11}(x)=
\begin{cases}
b_1, & 0\leq x \leq \frac{1}{2}a \\
0, & \frac{1}{2}a<x<L_{11}-\frac{1}{2}a\\
b_2, & L_{11}-\frac{1}{2}a \leq x \leq L_{11}
\end{cases}
\end{eqnarray}
\begin{eqnarray}
q_{12}(x)=
\begin{cases}
b_1, & 0\leq x \leq \frac{1}{2}a \\
0, & \frac{1}{2}a<x<L_{12}-\frac{1}{2}a\\
b_3, & L_{12}-\frac{1}{2}a \leq x \leq L_{12}
\end{cases}
\end{eqnarray}

\begin{eqnarray}
q_{1ij}(x)=
\begin{cases}
b_{i+1}, & 0\leq x \leq \frac{1}{2}a \\
0, & \frac{1}{2}a<x<\infty
\end{cases}
\end{eqnarray}
where $i,j=1,2$, $L_{11}$ and $L_{12}$ are lengths of bonds
$e_{11}$ and $e_{12}$ respectively.

Then for the generated soliton number we have
\begin{eqnarray}
N=\left<\frac{1}{2}+\frac{F_1+F_2+F_3}{\pi}\right>, \lab{number03}
\end{eqnarray}
where
\begin{eqnarray}
F_1=\frac{ab_1}{2}\left[
\sqrt{\frac{2}{\beta_1}}+\sqrt{\frac{2}{\beta_{11}}}+\sqrt{\frac{2}{\beta_{12}}}
\right]
\end{eqnarray}

\begin{eqnarray}
F_2=\frac{ab_2}{2}\left[
\sqrt{\frac{2}{\beta_{11}}}+\sqrt{\frac{2}{\beta_{111}}}+\sqrt{\frac{2}{\beta_{112}}}
\right]
\end{eqnarray}

\begin{eqnarray}
F_3=\frac{ab_3}{2}\left[
\sqrt{\frac{2}{\beta_{12}}}+\sqrt{\frac{2}{\beta_{121}}}+\sqrt{\frac{2}{\beta_{122}}}
\right]
\end{eqnarray}

The number of parameters in Eq.\re{number03} is much higher that
in the case of star graph. This implies that tree-shaped optical
fiber network provides more wider possibility for tuning the
generated soliton number and their dynamics.

\section{Conclusions}
In this paper we studied the problem of soliton generation in
optical fiber networks using a model based NLS equation on metric
graphs. Initial value (Cauchy) problem for NLS equation on metric
graphs is solved for different graph topologies, such as star,
tree and H-graphs. For branched optical fibers one can choose the
initial pulse profile in different ways (e.g., at the vertex or
branch, at given vertex or branch, with different shapes at
different vertices). Therefore, unlike to linear (unbranched)
fibers, soliton generation for optical fiber networks have richer
dynamics and tools for manipulation by solitons numbers. The above
method can be applied for different network topologies, provided a
network has three and more semi-infinite outgoing branches. This
allows to use our model for the problem of tunable soliton
generation in optical fiber networks, which is of importance for
practical applications in the areas, where optical fibers are used
for information (signal) transfer.

\section{Acknowledgements}
This work is supported by the grant  of the Ministry for
Innovation Development of Uzbekistan (Ref. No. BF2-022).

\end{document}